\documentclass[preprint,prb]{revtex4}
\usepackage{epsfig,graphics,bm}

\begin{document}

\title{Measurement of miniband parameters of a doped superlattice by photoluminescence
in high magnetic fields}
\author{R.F. Oliveira, A.B. Henriques, T.E.Lamas, A.A. Quivy,}
\affiliation{Instituto de F\'{\i}sica, Universidade de S\~ao Paulo\\
Caixa Postal 66318, 05315-970 S\~ao~Paulo, Brazil}
\author{E.Abramof}
\affiliation{LAS-INPE, Av. dos Astronautas, 1758 - Jd. Granja,
12227-010, S\~ao Jos\'e dos Campos, Brazil}
\begin{abstract}
We have studied a 50/50\AA\ superlattice of GaAs/Al$_{0.21}$Ga$_{0.79}$As
composition, modulation-doped with Si, 
to produce $n=1.4\times 10^{12}$~cm$^{-2}$ electrons per superlattice period. 
The modulation-doping was tailored to avoid the formation of Tamm states, and
photoluminescence due to interband
transitions from extended superlattice states was detected. By studying the effects of a quantizing magnetic
field on the superlattice photoluminescence, the miniband energy width, the reduced effective mass of the 
electron-hole pair, and the band gap renormalization could be deduced.
\end{abstract}
\maketitle

\section{Introduction}
The direct measurement of characteristic parameters of low dimensional semiconductor systems,
from their optical spectra,
requires an experimental technique sensitive enough to detect the
singularities in the electronic density of states [1-4]. 
In superlattices, optical detection of the van Hove 
singularities associated with electronic minibands
has been accomplished
by measuring the absorption spectrum due to intraband transitions,
in the far infrared wavelength range \cite{helm}.
Interband optical methods, however, could not be used successfully
to provide a direct measurement of the miniband width, due to the formation of saddle point 
excitons at
the edges of the van-Hove singularities \cite{grahn}, with a binding energy which is larger for the $M_1$ exciton
than for the $M_0$ exciton, due to a negative electronic effective 
mass along the growth direction in the former case \cite{gossard}. 
In heavily doped superlattices,
exciton formation is suppressed, due to Coulomb screening and phase-space filling. With sufficient doping, the
Fermi level will lay above the energy at top of the electronic miniband, and luminescence from miniband states
below the Fermi level should be observed.
One drawback, however, is that often 
the PL of doped superlattices is completely
dominated by transitions between localized Tamm states, precluding 
the detection of interband transitions associated with extended miniband states \cite{henriques12}.
When the doping atoms are located in the inner barrier layers of the superlattice, the spatial separation between the
electronic charge and the ionized donors gives rise to a strong bending of the band-edges at the
boundaries of the superlattice, which causes a shift of the outer wells from resonance with the inner ones, 
and Tamm states are formed (for a survey on Tamm states in superlattices see Ref.\cite{steslicka}).

If the modulation-doping profile is tailored 
in order to avoid the formation of Tamm states, then it should be possible
to detect the miniband singularities by using interband photoluminescence (PL) spectroscopy. 
By solving the Schroedinger
and Poisson equations, we found the formation of Tamm states is avoided if, 
in addition to the inner barriers, doping atoms are also added to the outer layers of the
superlattice, with an areal concentration about half the value used for the inner ones.
In this work we report an investigation of a 
superlattice sample with such a doping profile, grown by molecular beam epitaxy,
by photoluminescence in high magnetic fields. 
Experiments showed that when the doping level is such that the miniband is fully populated, i.e. the Fermi 
level lies above the $M_1$ singularity, luminescence between Landau levels of the $M_0$ 
and $M_1$ valence and conduction 
band states can be detected. The miniband parameters can be obtained directly
from the oscillations of the photoluminescence intensity as a function of magnetic field,
which is described by a doublet of frequencies. 
By measuring the oscillation
frequencies as a function of the photon energy, we can estimate the miniband energy width, 
the electron-hole pair reduced mass for in-plane 
movement, and the effective superlattice band gap. The superlattice band gap
is lowered by the self-energy correction, arising from the electron exchange interaction, 
by an amount equal to 
the so-called {\em band gap renormalization}\cite{schmittrink89} (BGR); 
For a superlattice with $1.42\times 10^{12}$~cm$^{-2}$ electrons per period,
our results show that the band gap is lowered by 23.4~meV.
This value is intermediate between the band gap renormalization of 
52~meV and 21~meV, estimated for a strictly two-dimensional
or three-dimensional electron gas of equivalent density, respectively.

\section{Experimental}
The superlattice sample was grown by molecular beam epitaxy (MBE) and consisted of 20 GaAs quantum 
wells of 50 \AA\ thickness, separated by 19 Al$_{0.21}$Ga$_{0.79}$As inner barriers of thickness 50 \AA.
Measurement of low-angle X-ray reflectivity measurements confirmed the thickness of the layers.
The internal AlGaAs barriers were delta-doped with Si at their center, with an areal
density of $1.5\times10^{12}$ cm$^{-2}$. The outer AlGaAs layers were delta-doped with half of the
same areal concentration, at a distance of 25 \AA\ from the adjacent GaAs layer.
Shubnikov-de Haas (SdH) measurements of in-plane conductivity were made on approximately square samples, 
with contacts in the corners,
using currents of $\sim$200-400 $\mu$A. 
PL measurements were done in the Faraday geometry,
using optical fibers and
{\em in situ} miniature focusing optics. 
All measurements were done at 2K.

\section{Results and Discussion}

Figure 1 shows the Fourier transform of the SdH oscillations for the
superlattice sample, for a magnetic field applied normal to the surface of the sample.
Two peaks are observed, at $f_{SdH}(M_1)=23.4$~T and $f_{SdH}(M_0)=34.9$~T. These peaks are associated with the 
'belly' and 'neck' extremal orbits in the mini-Fermi surface \cite{stormer86}, 
and can be labeled by the
corresponding saddle points in the density of states, $M_0$ and $M_1$. 
To demonstrate this association, SdH measurements were done for a magnetic field direction tilted
away from the normal to the surface of the sample. In tilted fields, 
the two peaks show a characteristic behaviour of quasi-three dimensional
electrons, whereby the $M_0$ and $M_1$ peaks cross over \cite{iye,prb00}, as shown in Figure 2.
According to Onsager's quasiclassical quantization formula,  
the Fourier frequencies in the SdH oscillations will be given by 
$f_{\mbox{\scriptsize SdH}}=\frac{\hbar}{2\pi e}{\cal A}_e$, where ${\cal A}_e$ are the extremal sections
of the mini-Fermi surface, hence for a parabolic electronic in-plane dispersion we obtain
\begin{equation}
f_{\mbox{\scriptsize SdH}}(M_0)=\frac{m_e\phi_e}{\hbar e},\,\,\,\,
f_{\mbox{\scriptsize SdH}}(M_1)=\frac{m_e\left(\phi_e-\Delta_e\right)}{\hbar e},
\label{eq:onsager}
\end{equation}
where $m_e$ is the electronic in-plane effective mass, $\phi_e$ is the Fermi energy,
and $\Delta_e$ is the energy width of the miniband.
The full and dashed lines shown in figure 2 are the theoretical $M_0$ and $M_1$ SdH frequencies, 
respectively, obtained from equation (\ref{eq:onsager}), 
by solving self-consistently the Schroedinger and Poisson equations, for an infinite
Al$_{0.21}$Ga$_{0.79}$As/GaAs
superlattice, with all GaAs and Al$_{0.21}$Ga$_{0.79}$As layers 
being of width 50\AA, and doped in the middle of the internal barriers 
with $N_d=1.42\times 10^{12}$~cm$^{-2}$. 
The good agreement between theoretical and experimental Fourier frequencies
demonstrates the presence of electrons confined by a superlattice potential,
in close agreement with the design parameters. 

The electronic effective mass, $m_e$, was estimated
from the temperature dependence of the amplitude of the SdH oscillations, giving $m_e=0.068 m_0$.
Using the experimental values, $m_e=0.068 m_0$, $f_{SdH}(M_1)=23.4$~T and $f_{SdH}(M_2)=34.9$~T,
equation (\ref{eq:onsager}) gives $\Delta_e=19.6$~meV and $\phi_e=59.4$~meV.

Figure 3 shows the photoluminescence (PL) spectrum at $T=2$~K
for the superlattice. A broad emission band is seen above the GaAs gap, which is associated with
recombinations between electrons and holes confined by the superlattice potential.
The effective superlattice band gap, $E_0$, the energy corresponding to $E_0+\Delta_e$,
and the energy corresponding to $E_0+\phi_e$,
are shown in figure 3; these energies were estimated from the analysis below.

In order to demonstrate that the wide PL emission band shown in figure 3 arises from electron-hole
recombinations by charge carriers confined by the superlattice potential, the PL spectra were studied as a
function of the applied magnetic field. Figure 4 shows the PL intensity oscillations at fixed photon energies.
The oscillations in the PL intensity, at a given photon energy $h\nu$, 
will be proportional to \cite{bastard}
\begin{eqnarray}
I(h\nu,B)\sim\sum_{N,N',k_e,k_h}\left|\left<\phi_h(N',k_h)|\phi_e(N,k_e)\right>\right|^2
\times\nonumber\\
\times\delta\left[E_0+\left(N+\frac{1}{2}\right)\hbar\omega_e+E_e(k_e)+
\left(N'+\frac{1}{2}\right)\hbar\omega_h+E_h(k_h)-h\nu\right]
\end{eqnarray}
where the summation is over electronic states, $(N,k_e)$,
that are situated below the Fermi level, and over hole states, $(N',k_h)$, that contain
photoexcited holes. Due to momentum conservation \cite{roth,voisin}, transitions occur only if $N'=N$ and $k_h=k_e=k$,
and we obtain
\begin{equation}
I(h\nu,B)\sim\left|\left<\chi_h|\chi_e\right>\right|^2\sum_{N,k}
\delta\left[E_0+\left(N+\frac{1}{2}\right)\frac{\hbar eB}{\mu}+E_e(k)+E_h(k)-h\nu\right]
\label{eq:interm}
\end{equation}
where $\chi_e$, $\chi_h$ are the electron and hole superlattice envelope wave functions, and $\mu$ is the
reduced mass of the electron-hole pair.
In the tight-binding approximation the electronic and hole miniband dispersion will be given by, 
\begin{equation}
E_e(k)=\frac{\Delta_e}{2}(1-\cos kd),\hspace{1cm} E_h(k)=\frac{\Delta_h}{2}(1-\cos kd),
\label{eq:tb}
\end{equation}
where $d$ is the period of the superlattice,
and by following the steps described in \cite{prb94}, from equations (\ref{eq:interm},\ref{eq:tb})
we can obtain an analytical expression for the PL intensity
oscillations:
\begin{equation}
I(h\nu,B)\sim -\exp\left(-\frac{2\pi\gamma\mu}{\hbar e B}\right)
J_0\left(\pi
\frac{\Delta_e+\Delta_h}{\hbar eB/\mu}
\right)
\cos\left(2\pi\frac{h\nu-E_0-\frac{\Delta_e+\Delta_h}{2}}{\hbar eB/\mu}\right)
\label{eq:plosc}
\end{equation}
where $\gamma$ is an energy level broadening parameter. Equation (\ref{eq:plosc}) yields a doublet of
frequencies of oscillation, associated with saddle points $M_0$ and $M_1$, whose values are

\begin{equation}
f_{PL}(M_0)=\mu\frac{h\nu-E_0}{\hbar e},\hspace{10mm}f_{PL}(M_1)=\mu\frac{h\nu-E_0-\Delta_e-\Delta_h
}{\hbar e}
\label{eq:freq}
\end{equation}

Figure 5 shows the oscillations in the PL intensity as a function of magnetic field for a photon energy
of $h\nu=1.6199$~eV, and a theoretical fit to equation (\ref{eq:plosc}), including a monotonous parabolic background.
The fitting procedure yields the values for $f_{PL}(M_0)=22.2$~T, and $f_{PL}(M_1)=8.0$~T. 

The fitting procedure was repeated for all photon energies measured.
Figure 6 shows the frequencies of oscillations of the PL intensity in a magnetic field as 
a function of the photon energy. A linear dependence of the PL oscillation frequencies
on the photon energy is obtained, in agreement with eq.(\ref{eq:freq}). 

The straight lines depicted 
in figure~6 were obtained from a simultaneous linear fit of the two sets of data points with eq.(\ref{eq:freq}).
The fit yields the intersection of the lines with the energy axis, which according to eq.(\ref{eq:freq})
will occur at the energies $h\nu=E_0=1.581$~eV and $h\nu=E_0+\Delta_e+\Delta_h=1.604$~eV.
The heavy-hole miniband energy width can be estimated theoretically from the solution of 
the Schroedinger-Poisson equations for the given structure, and we obtain $\Delta_h=0.3$~meV, i.e. the
hole dispersion can be ignored as a good approximation, hence we can deduce
$E_0=1.581$~eV and 
$\Delta_e=23$~meV. The latter parameter is in reasonable agreement with the electronic miniband width estimated 
from the Shubnikov-de Haas measurements, $\Delta_e=19.6$~meV. The slope of the line determines the 
reduced effective mass of the electron-hole pair, i.e. $\mu=0.0612\,m_0$.
The estimated superlattice band gap, $E_0=1.581$~eV,
can be compared to the theoretical value of the band gap, obtained from the $\bm k\cdot \bm p$
equation for this structure, $E_0^{th}=1.605$~eV, which does not take into account 
the lowering of the band gap due to many body effects. By equating $E_0=E_0^{th}+$BGR,
we obtain BGR$=-24$~meV.
For an electron gas confined in two dimensions, the band gap renormalization is \cite{sr_ssc84}
BGR$^{2D}=-3.1(n a_X^2)^{1/3}E_X$~meV, where $a_X$ and $E_X$ are the exciton Bohr radius and
binding energy, respectively, and $n=1.42\times 10^{12}$~cm$^{-2}$ is the areal density of electrons
in the well.
For a strictly two-dimensional electron gas, we substitute $E_X=4R^*$, 
and an exciton radius of $a_X=a_B^*/2$, where $R^*$ and $a_B^*$ are the effective Rydberg
and effective Bohr radius in bulk GaAs, respectively, to obtain BGR$^{2D}$=52~meV.
The finite thickness of the charges can be taken into account by calculating the exciton radius and binding energy
for a 50\AA\ GaAs/Al$_{0.21}$As$_{0.79}$As quantum well, as described in Refs \cite{dug90,slm92}, 
to give $E_X=10.34$~meV and $a_X=53$~\AA, in which case
BGR$^{Q2D}=-28$~meV. Finally, in the bulk the BGR is given by \cite{reinecke_prl_79}
BGR$=-3.5\left[n_{3D}(a_B^*)^3\right]^{1/4}$, where $n_{3D}=n/d$ is the equivalent bulk carrier concentration,
which gives BGR$^{3D}=-21.9$~meV. Thus, we find a BGR that is
slightly larger than in the bulk, but is less than the BGR for a quasi two-dimensional system.

In conclusion, superlattice structures  of GaAs/AlGaAs composition were produced, 
with a modulation doping profile that prevents the formation of Tamm states.
Shubnikov-de Haas measurements in tilted fields demonstrate the presence of 
electrons confined by a superlattice potential in close agreement with the design parameters. 
PL was studied in high magnetic fields, and a broad luminescence band was observed above the
GaAs band gap. For a fixed photon energy within this broad PL band, 
the intensity of the luminescence oscillates as a function of the magnetic field applied perpendicular
to the superlattice layers. The PL oscillations are described by a doublet of frequencies, which show
a linear dependence on the energy of the photon, showing that it is due to
recombinations of electron-hole pairs confined by the superlattice
potential. From the photoluminescence oscillations we can estimate superlattice parameters - the electronic
miniband width $\Delta_e$, the electron-hole pair reduced mass, $\mu$, and the band gap renormalization,
which we find 
is less than the BGR for a quantum well of the same barrier composition and well thickness, but greater than 
the BGR for
a bulk GaAs degenerate electron gas.

\section{Acknowledgments}
This work was supported by FAPESP (Grants No. and 99/10359-7 and No. 01/00150-5) and 
CNPq (Grant No. 306335/88-3 ).

\pagebreak
\begin{minipage}[b]{5.6in}
\section*{FIGURE CAPTIONS}
{\bf Fig.~1} Fourier transform of the SdH oscillations for sample 2268, for a magnetic field
applied normal to the surface of the sample. The labels indicate the saddle point in the density of states
that originates a given frequency peak.\\ \\

{\bf Fig.~2} Peak position of the Fourier transform of the SdH
oscillations as a function of the angle between the magnetic field direction and the normal to the surface of
the sample. The area of each dot is proportional to the intensity of the peak. The full and dashed lines
correspond to the theoretical frequencies 
for the oscillatory components associated with saddle points $M_0$ and
$M_1$, respectively.\\ \\

{\bf Fig.~3} PL spectrum for the AlGaAs/GaAs superlattice. The superlattice band gap is indicated.\\ \\

{\bf Fig.~4} PL intensity oscillations at fixed photon energies. The photon energy corresponding to
each curve is shown.\\ \\

{\bf Fig.~5} PL intensity oscillations at a photon energy of $h\nu=1.6199$~eV. Dots are the experimental values
and the full line is a fit with equation (\ref{eq:plosc}). The best fit values of the adjusting parameters 
are shown.\\ \\

 {\bf Fig.~6} PL intensity oscillations at a photon energy of $h\nu=1.6199$~eV. Dots are the experimental values
and the full line is a fit with equation (\ref{eq:plosc}). The best fit values of the adjusting parameters 
are shown.\\ \\

\end{minipage}

\pagebreak
\setlength{\unitlength}{1mm}
\begin{figure}[h]
\centerline{\hspace{-1cm}\epsfxsize=8cm\epsffile{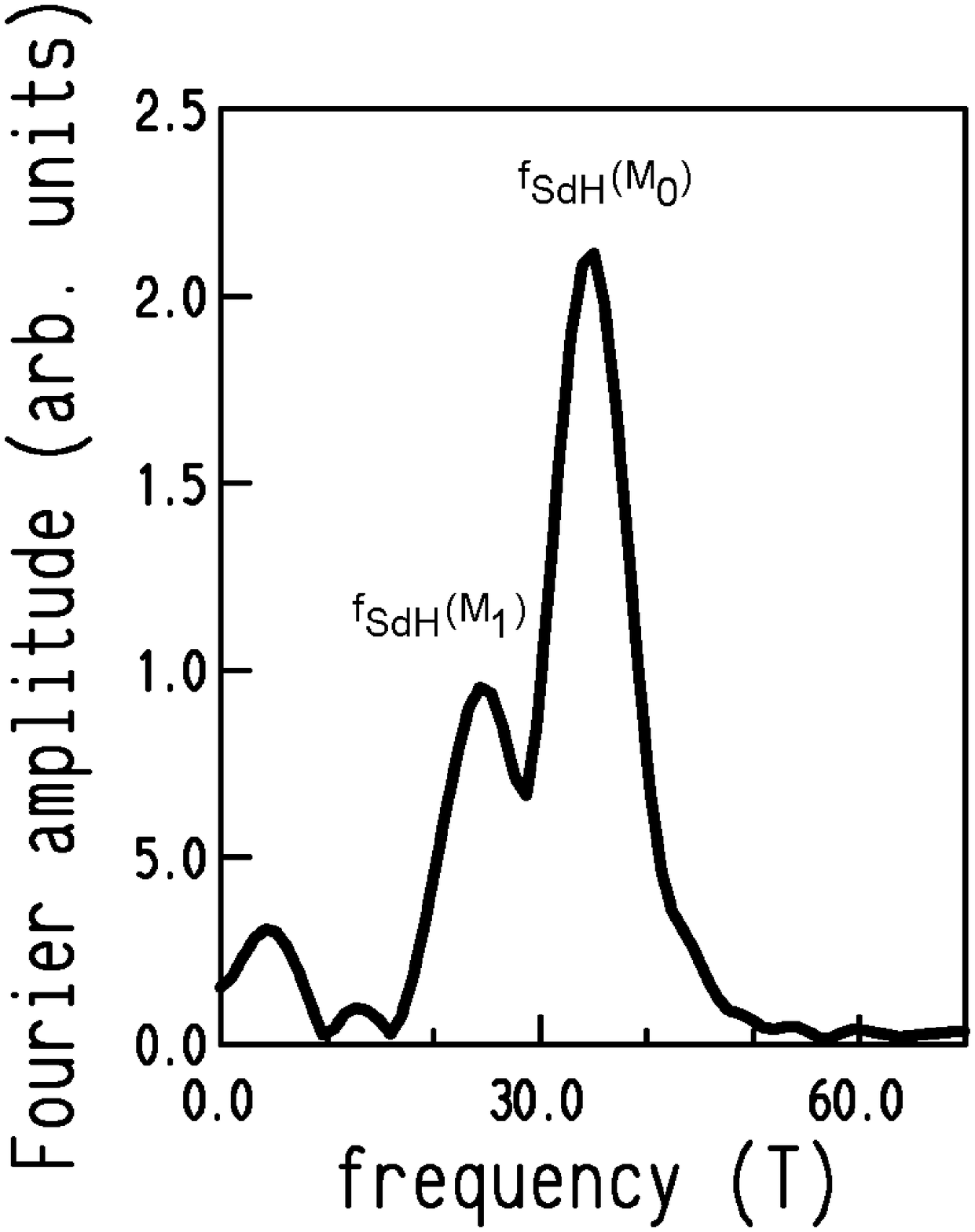}}
\caption{}
\label{fig:fig1}
\end{figure}
\vfill
{\tt Manuscript by Henriques {\em et al}, Figure 1}

\pagebreak
\setlength{\unitlength}{1mm}
\begin{figure}[h]
\centerline{\hspace{-1cm}\epsfxsize=8cm\epsffile{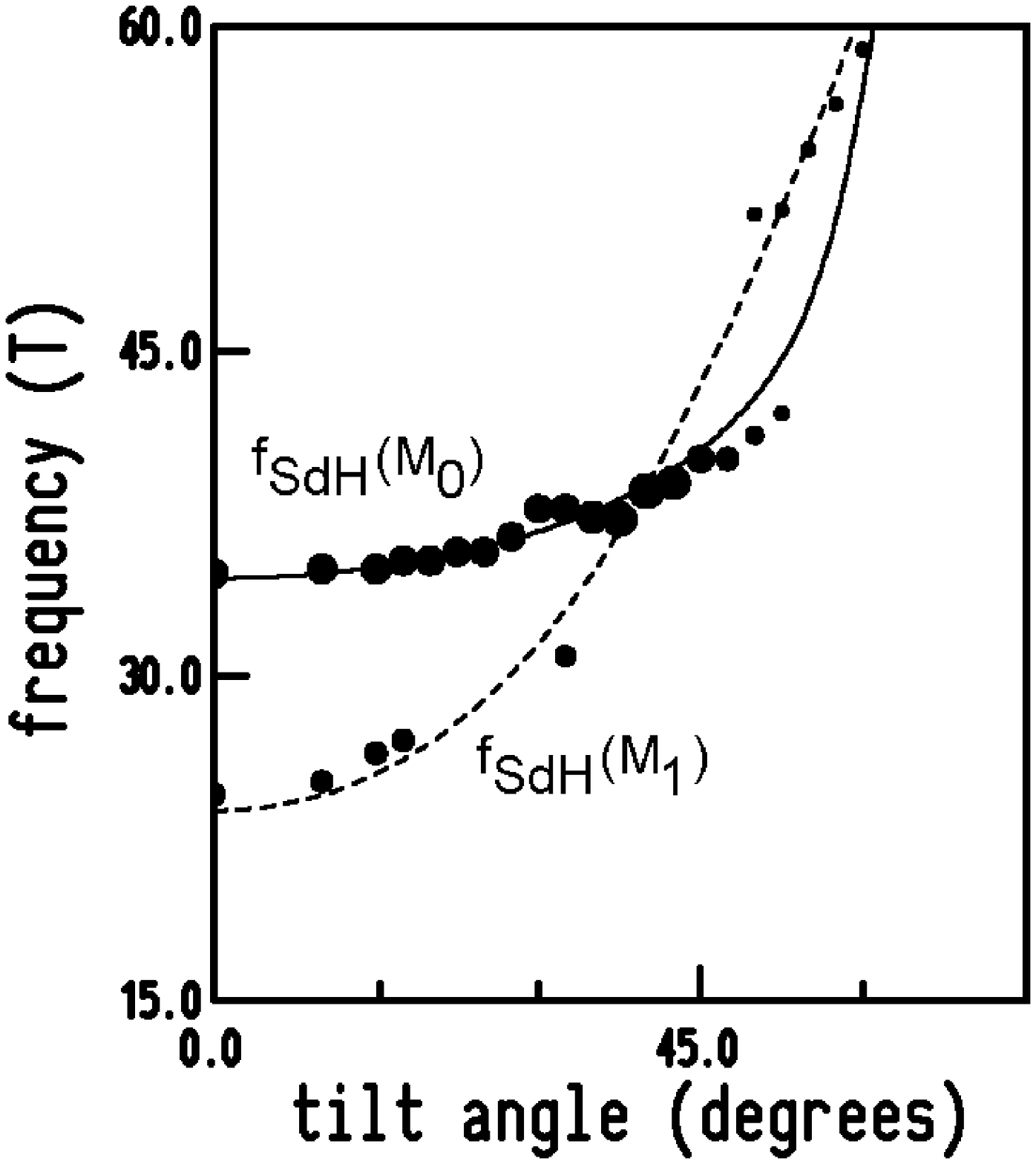}}
\caption{}
\label{fig:fig2}
\end{figure}
\vfill
{\tt Manuscript by Henriques {\em et al}, Figure 2}

\pagebreak
\setlength{\unitlength}{1mm}
\begin{figure}[h]
\centerline{\hspace{-1cm}\epsfxsize=8cm\epsffile{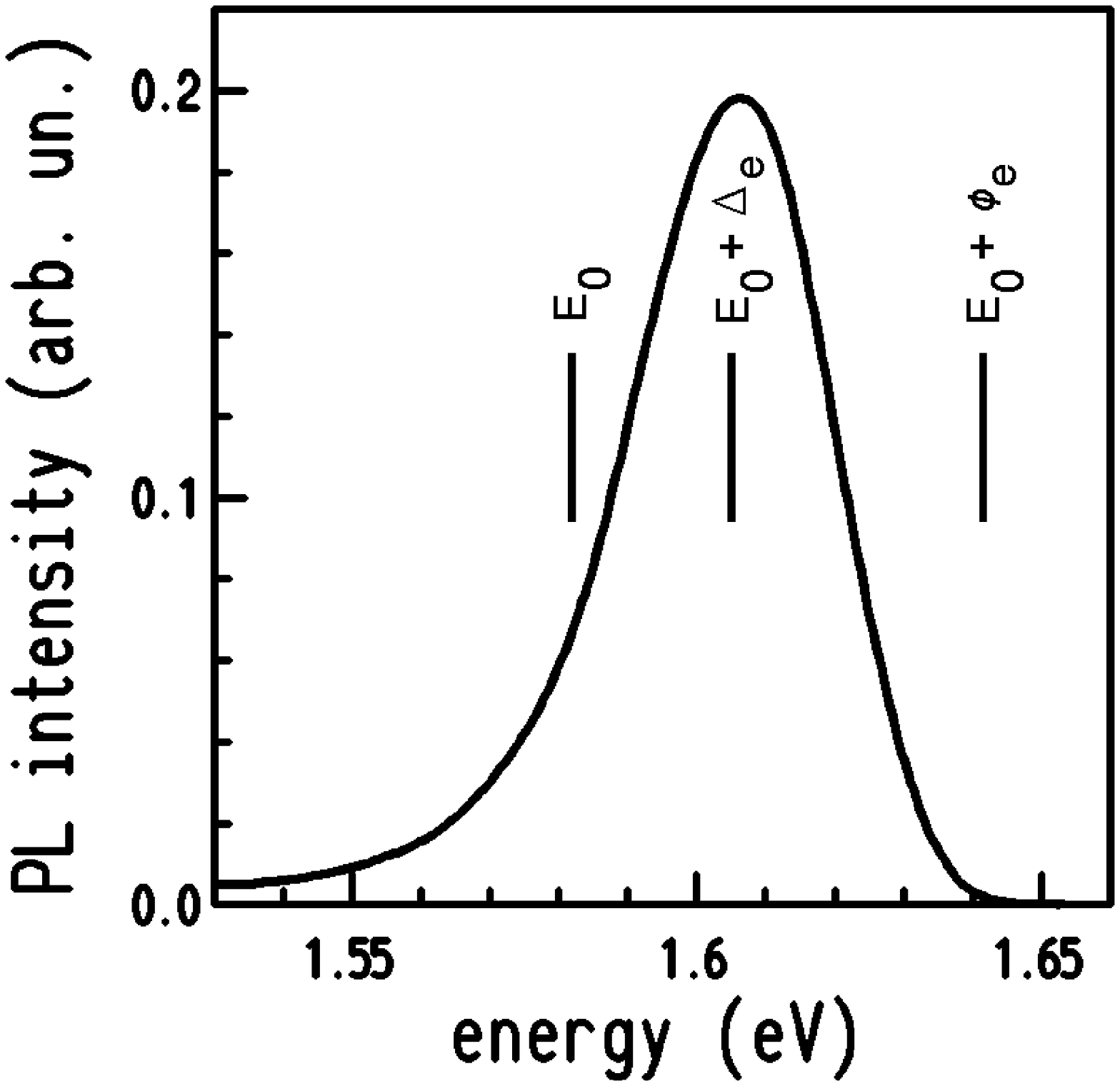}}
\caption{}
\label{fig:fig3}
\end{figure}
\vfill
{\tt Manuscript by Henriques {\em et al}, Figure 3}

\pagebreak
\setlength{\unitlength}{1mm}
\begin{figure}[h]
\centerline{\hspace{-1cm}\epsfxsize=8cm\epsffile{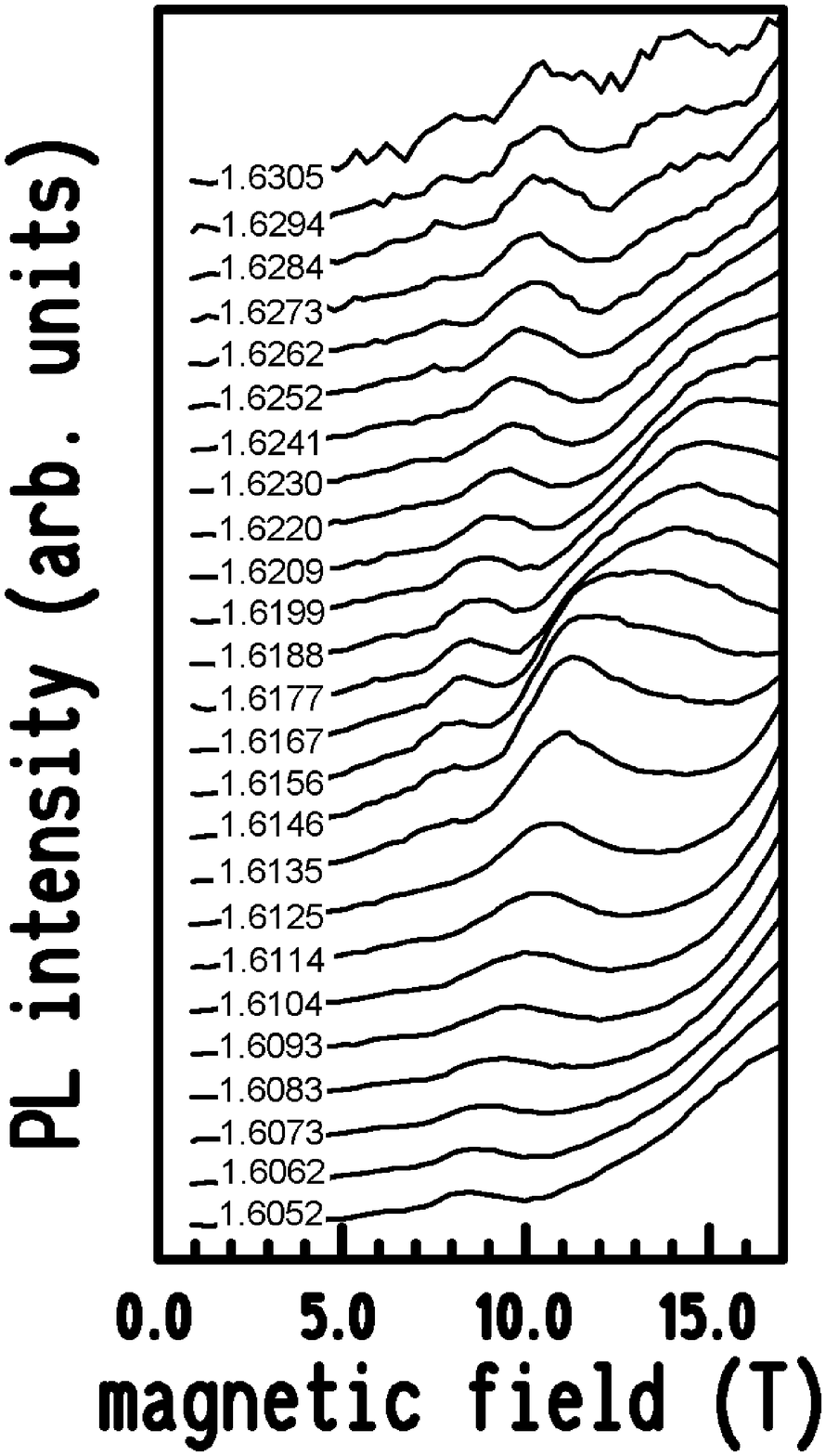}}
\caption{}
\label{fig:fig4}
\end{figure}
\vfill
{\tt Manuscript by Henriques {\em et al}, Figure 4}

\pagebreak
\setlength{\unitlength}{1mm}
\begin{figure}[h]
\centerline{\hspace{-1cm}\epsfxsize=8cm\epsffile{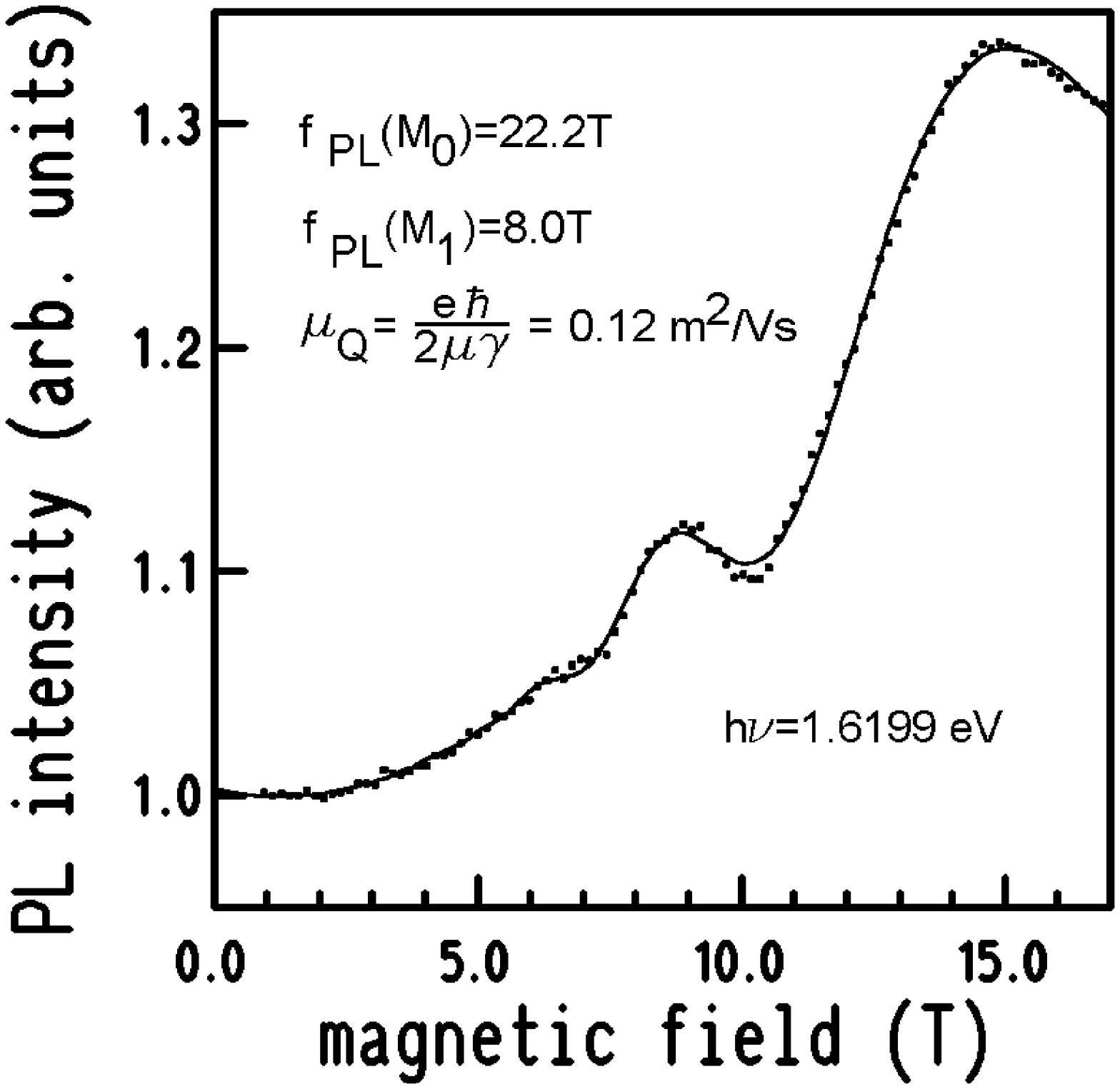}}
\caption{}
\label{fig:fig5}
\end{figure}
\vfill
{\tt Manuscript by Henriques {\em et al}, Figure 5}

\pagebreak
\setlength{\unitlength}{1mm}
\begin{figure}[h]
\centerline{\hspace{-1cm}\epsfxsize=8cm\epsffile{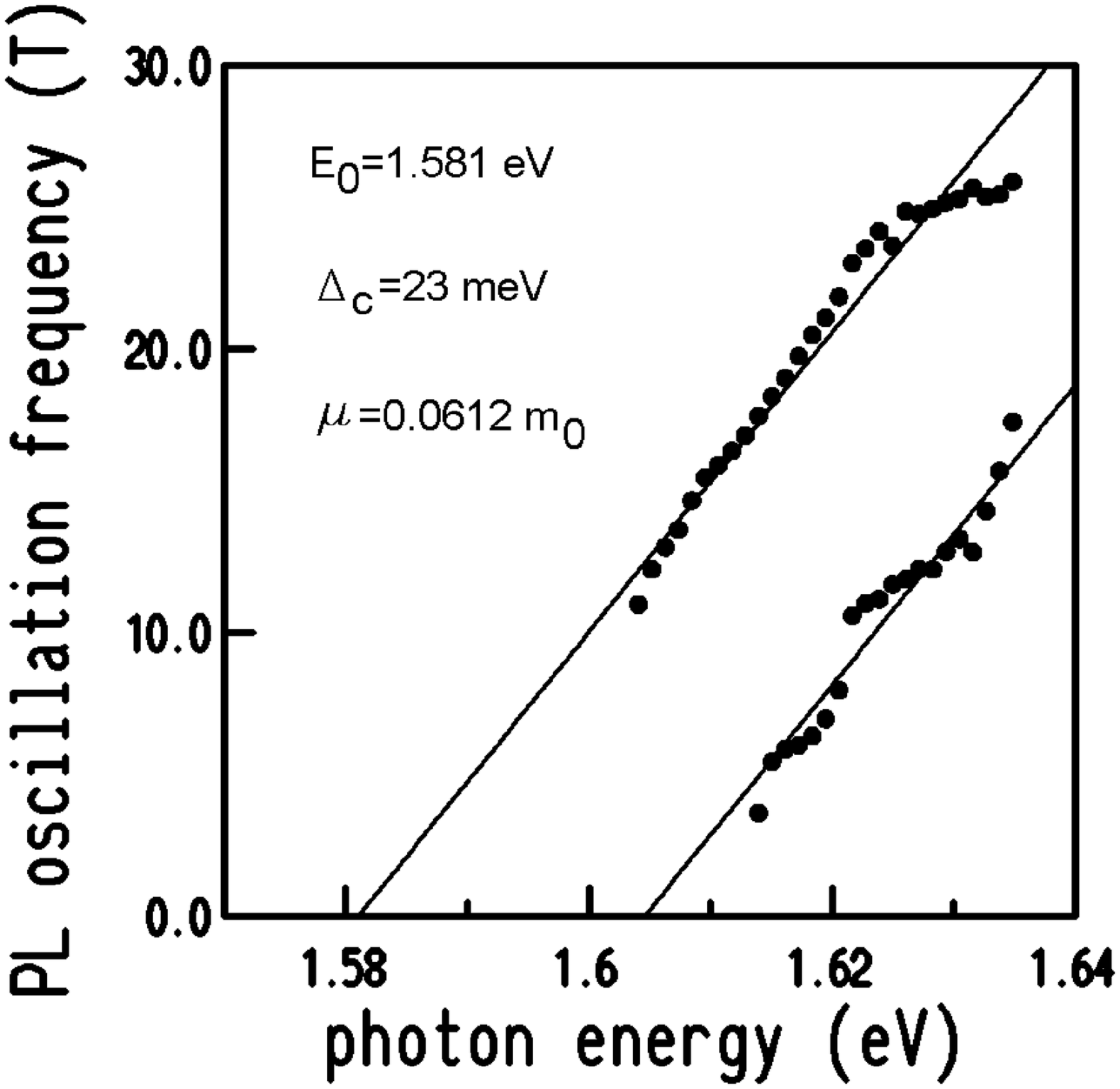}}
\caption{}
\label{fig:fig6}
\end{figure}
\vfill
{\tt Manuscript by Henriques {\em et al}, Figure 6}

\end{document}